\documentclass[useAMS,usenatbib]{mn2e}
\usepackage[utf8]{inputenc}
\usepackage{amsmath}
\usepackage{wasysym}
\usepackage{amsfonts}
\usepackage{amssymb}
\usepackage{graphicx}
\usepackage{color}
\usepackage{xspace}

\title{Using failed supernovae to constrain the Galactic r-process element production}
\author[Wehmeyer et al.]{B. Wehmeyer,$^{1,6}$ \thanks{bwehmey@ncsu.edu} C. Fr\"ohlich,$^{1,6}$ B. C\^ot{\'e},$^{2,6,7}$ M. Pignatari,$^{3,2,6,7}$ and F.-K. Thielemann$^{4,5}$ \\
$^1$Department of Physics, North Carolina State University, 2401 Stinson Dr, Raleigh, NC 27695-8202, USA\\
$^2$Konkoly Observatory, Research Centre for Astronomy and Earth Sciences,\\ Hungarian Academy of Sciences, Konkoly-Thege Miklós út 15-17, H-1121 Budapest, Hungary\\
$^3$E.A. Milne Centre for Astrophysics, Dept. of Physics \& Mathematics, University of Hull, HU6 7RX, United Kingdom\\
$^4$Univ. Basel, Dept. Phys., Klingelbergstr. 82, CH-4056 Basel, Switzerland\\
$^5$ GSI Helmholtzzentrum f\"ur Schwerionenforschung, Planckstraße 1, 64291 Darmstadt, Germany\\
$^6$Joint Institute for Nuclear Astrophysics - Center for the Evolution of the Elements\\
$^7$NuGrid Collaboration, http://nugridstars.org\\
}
\date{Accepted 2019 May 05. Published 2019 May 16; in original form 2018 December 31}
\begin{document}
\maketitle
\begin{abstract}

Rapid neutron capture process (r-process) elements have been detected in a large fraction of metal-poor halo stars, with abundances relative to iron (Fe) that vary by over two orders of magnitude. This scatter is reduced to less than a factor of three in younger Galactic disk stars. The large scatter of r-process elements in the early Galaxy suggests that the r-process is made by rare events, like binary compact mergers and rare sub-classes of supernovae. Although being rare, neutron star mergers alone have difficulties to explain the observed enhancement of r-process elements in the lowest-metallicity stars compared to Fe. The supernovae producing the two neutron stars already provide a substantial Fe abundance where the r-process ejecta from the merger would be injected. In this work we investigate another complementary scenario, where the r-process occurs in neutron star - black hole mergers in addition to neutron star mergers. Neutron star - black hole mergers would eject similar amounts of r-process matter as neutron star mergers, but only the neutron star progenitor would have produced Fe. Furthermore, a reduced efficiency of Fe production from single stars significantly alters the age-metallicity relation, which shifts the onset of r-process production to lower metallicities. We use the high resolution ((20 pc)$^3$ /cell) inhomogeneous chemical evolution tool “ICE” to study the outcomes of these effects. In our simulations, an adequate combination of neutron star mergers and neutron star-black hole mergers qualitatively reproduces the observed r-process abundances in the Galaxy.

\end{abstract}

\section {Introduction}\label{introduction}
The $r$-process (e.g., \citealt{Cowan91,Arnould07,Thielemann11,Cowan19}, and references therein) is one of the dominant sources of elements heavier than Fe. At present, it is still unclear whether neutron star mergers (NSMs, since recently the only observed and confirmed r-process site) are the \textit{exclusive} site of this process (e.g., \citealt{Cescutti15,Hirai15,Ishimaru15,Shen15,vandeVoort15,Wehmeyer15,Thielemann17,Cote18,Hotokezaka18,Ojima18,Siegel18,Cowan19,Haynes19}).
While early scientific studies argued that neutrino fluxes in core-collapse supernovae (CCSNe) would have the right properties to host neutrino-driven nucleosynthesis (e.g., \citealt{Arcones13}, and references therein) which might include the r-process (e.g., \citealt{Woosley94}, \citealt{Takahashi94}), later and more advanced calculations (e.g., \citealt{Liebendorfer03}) pointed to proton-rich conditions in their innermost ejecta, rather causing a $\nu p$ process (\citealt{Frohlich06a,Frohlich06b,Pruet05,Pruet06,Wanajo06,Wanajo13}) instead of the r-process.
However, the collapse of the core of a massive star leads either to a CCSN and a neutron star (NS) or the formation of a black hole (BH, e.g., \citealt{Heger03}).
When two NSs merge (e.g., \citealt{Lattimer74,Paczynski86,Eichler89}, but see also more recent works, e.g., \citealt{Rosswog18}), conditions for the onset of the $r$-process are met (\citealt{Freiburghaus99,Panov08,Korobkin12,Bauswein13,Rosswog13,Rosswog14,Wanajo14,Eichler15,Just15}). This site has been confirmed 
by gravitational wave detection GW170817 (e.g., \citealt{Abbott1}), followed by its optical counterpart, kilonova SSS17a, showing evidence of the successful production of $r$-process elements (e.g., \citealt{Abbott2}, \citealt{Abbott3}). Hence, NSMs are a confirmed source of Galactic $r$-process elements. 
Considering this site as \textit{the exclusive} $r$-process site, however, comes with two distinct issues:
\begin{enumerate}
    \item \textit{$r$-process elements are abundant already at very low metallicities.} 
    Two CCSNe must have occurred before the NSM event in order to produce the two involved NSs. Hence, the interstellar medium (ISM) hosting the NSM is already polluted by the Fe-rich ejecta of those two CCSNe. 
    Many stars with low metallicity already show high $r$-process abundances compared to Fe, up to two orders of magnitude larger than solar (e.g., \citealt{Sneden08,Roederer10,Hansen18}).
     Such enhancements are difficult to explain by a scenario where NSM act as \textit{the exclusive} r-process element production source (e.g., \citealt{Argast04,Wehmeyer15}).
    \item \textit{$r$-process elemental abundances in low metallicity stars show a large scatter in comparison to solar metallicity stars.}
    The observed abundance scatter in alpha elements\footnote{Among the stable alpha elements are C, O, Ne, Mg, Si, S, Ar, and Ca.} with respect to Fe remains rather small throughout the entire chemical evolution. Instead, $r$-process elements show a much larger scatter in abundances at low metallicities (\citealt{Roederer10}, \citealt{Beers18}).  Since alpha elements are made mostly by CCSNe, this suggests that $r$-process elemental production events should occur at a lower frequency than CCSNe (e.g., \citealt{Thielemann17}).

\end{enumerate}
Recent works to address these open questions have mostly considered two scenarios, i.e., adding a rare sub-class of supernova as second early r-process site, or considering sub-halos of the Galaxy as independent building blocks that will later merge to form the Galaxy.
The former approach is based on the assumption that there could be a second, rare $r$-process production site, e.g., (the sub-class of) magnetorotationally driven supernovae (or ``jet-supernovae'', see, e.g., \citealt{Winteler12,Mosta15,Nishimura17,Halewi18}). Since this site would eject $r$-process elements and negligible amounts of Fe, $r$-process elements could be released into a region of lower metal content than a NSM could (those require two NSs to be present and thus two previous supernovae, already enhancing the ISM with metals), if the occurrence rate of such a supernova would be low (as expected due to the required high magnetic fields) in comparison to ``regular'' CCSNe. Stars being polluted by such an event would inherit high $r$-process abundances in comparison to stars polluted by regular CCSNe. This would also allow to explain the large scatter seen in $r$-process abundances in low metallicity stars. 

These considerations were already discussed in \cite{Cescutti15} and \cite{Wehmeyer15}. However, despite the fact that $10^{15}$ Gauss NSs as remnants of this distinct supernova channels have been detected, this scenario still has to wait for an observational confirmation (\citealt{Fujimoto06}, \citealt{Fujimoto08}, \citealt{Winteler12}, \citealt{Mosta14}). Furthermore, this nucleosynthesis site involves the difficulty of high resolution treatments of the magneto-rotational instability (e.g., \citealt{Mosta15,Rembiasz16,	Sawai16,Nishimura17,Obergaulinger18}).

A second approach to solve the difficulties (i) and (ii) above, is to consider dwarf galaxies as individually developing sub-systems that will merge to later form the Galactic halo (e.g., \citealt{Hirai15}).
Observations of dwarf galaxy systems show that these systems have lower star formation efficiency (\citealt{Kirby13}) and higher gas outflow rates (see predictions from cosmological simulations, e.g., \citealt{Muratov15,Pillepich18}). 
These features allow the contribution of NSMs to already take place at low metallicities (because lower star formation efficiency slows down the temporal evolution of [Fe/H], which allows NSMs to appear at lower [Fe/H] values with respect to the star formation rate, cf., \citealt{Ishimaru15}) and provide large abundance scatter (among others, because of gas outflows in chemodynamical models, cf., \citealt{Hirai15}, and the stochastic nature of dwarf systems, cf., \citealt{Ojima18}).
Although such systems are observationally confirmed to have seen $r$-process production events (\citealt{Ji16,Marshall18}), it is yet unclear whether a stochastic chemical evolution model featuring low star formation efficiencies is applicable to the bulk of these kind of systems (\citealt{Kirby13,Ojima18}).

In this paper, we study an alternative  scenario with respect to the ones discussed above: We consider BH -  NS mergers (BHNSM) as second $r$-process elemental production site in addition to NSMs. 
This site has one major difference compared to NSMs: BHNSM require only \textit{one} NS to be present in the system. This means that only \textit{one} CCSN is required in the system before the $r$-process event. This allows BHNSMs to occur at lower initial metallicities than NSMs. Also, the slower overall increase of metallicity due to less successful CCSNe permits the presence of r-process rich stars at lower metallicities.

This work is organized as follows. In section~\ref{Observations}, we discuss the astronomical observations relevant for this work. In section~\ref{The model}, we introduce the model used to compute the evolution of abundances. In section~\ref{Results}, we present the influence of the different r- and non-r-process sites on the evolution. In section~\ref{Conclusions and Discussion}, our results are summarized and discussed.

\section {Observations}\label{Observations}

\subsection{Europium as tracer of r-process elements}

Galactic chemical evolution (GCE) is a powerful tool to study the contributions of the different elemental production sites to stellar abundances.
For many lighter elements (e.g., Mg, O, C) the production sites are well known. Beyond Fe, the r-process contributions provide about half of the element abundances in the solar system, and are the dominant source in the Universe of several elements like Ir, Pt and Au (for a recent review see \citealt{Cowan19}). Eu is the most observed r-process element, and it is used as a diagnostic to study the history of the r-process enrichment of the Galaxy (e.g., \citealt{Burris00}). Eu abundances are derived using mostly the two UV lines at $4192.70$ and $4205.05$ Angstr\"om (e.g., \citealt{Biemont82}).

We make use the abundance database SAGA (Stellar Abundances for Galactic Archaeology, e.g., \citealt{Suda08,Suda11,Yamada13}), with [Eu/Fe]\footnote{We use the notation [A/B]$=\log (\text{A}/\text{B})_\text{star}-\log (\text{A}/\text{B})_\odot$} abundances mainly from \cite{Francois07,Simmerer04,Barklem05,Ren12,Roederer10,Roederer14a,Roederer14b,Roederer14c,Shetrone01,Shetrone03,Geisler05,Cohen09,Letarte10,Starkenburg13,McWilliam13}. We exclude carbon enhanced metal poor (``CEMP'') stars i.e., stars with [C/Fe]$\geq 1$ and [Fe/H]$\leq -1$) and stars with binary nature, since the surface abundances of such objects are expected to be affected by pollution from a binary companion (\citealt{Ryan2005}), which is beyond the scope of the present study. 
When comparing the observed Eu abundances as a function of [Fe/H] with those of  lighter alpha elements (primarily those made by CCSNe) it is very striking to see that the two curves behave similarly close to solar metallicities, but differ greatly at low metallicities (e.g., \citealt{Thielemann17,Cowan19}), making metal-poor stars to unique tracers of the early evolution of Galactic r-process nucleosynthesis (e.g., \citealt{Sneden08,Frebel18,Horowitz18}).

\subsection{GW170817/SSS17a}
The detection of the gravitational wave event GW170817 (e.g., \citealt{Abbott1}) has been interpreted as a coalescence of two compact objects with masses in the range $1.17 \text{M}_{\odot} \leq m \leq 1.60 \text{M}_{\odot}$. The GW emission was followed by the detection of a kilonova (SSS17a) whose light curve suggests r-process element production (e.g., \citealt{Chornock17,Cowperthwaite17,Tanaka17,Villar17}).
Lanthanides as Eu were produced in the event (e.g., \citealt{Tanvir17,Wollaeger18}). While the majority of the literature suggests that the coalescence of two NSs was the origin of this astronomical event (\citealt{Abbott4}), it cannot be ruled out that the event was actually the coalescence of a NS and a BH (\citealt{Hinderer18}). Furthermore, \cite{Hinderer18} showed that the \textit{GW only} and the \textit{electromagnetic only} observations can only rule out a BHNSM for an extreme range of the parameter space and find that 40\% of the parameter space set by the nuclear and astrophysical uncertainties would permit a BHNSM event instead of a NSM event in the case of GW170817/SSS17a. 
A possible formation channel for a required stellar mass BH - considered in this study - is that it originates in a failed SN (e.g., \citealt{Heger03}), which will be discussed in sections~\ref{model:SNe}~{\&}~\ref{subsec:NSM_BHNSM}.
Another possible origin of the required stellar mass BH is e.g., in primordial fluctuations in the early Universe. A probable formation channel of such objects is described in e.g., \cite{Garcia96,Carr16,Garcia18}. However, their occurrence frequency in BHNSMs is hard to predict, therefore we do not include them here explicitly.

\section {The GCE model}\label{The model}
In comparison to homogeneous GCE models, inhomogeneous models track the \textit{location} of the nucleosynthesis sites. This permits to reproduce the \textit{scatter} of abundances instead of predicting a linear evolution. On the other hand, large scale effects (e.g., galaxy collisions, spiral arms mixing) can only be approximated in such models.
In this study we use the inhomogeneous chemical evolution model described in \cite{Wehmeyer15}. In the following sections, we recall the main components of the model for convenience (sections~\ref{model:setup}, \ref{model:LIMS}, and \ref{model:SNIa}) and highlight the improvements made to the model for the purpose of this study, especially the treatment of the additional r-process site related to BHNSMs (sections~\ref{model:SNe} and \ref{subsec:NSM_BHNSM}).
\subsection{General setup}\label{model:setup}

We set up a cube of ($2$ kpc)$^3$ in the Galaxy which is cut into $100^3$ sub cubes with an edge length of 20 pc.  
During each time step of 1 My, the following calculations are performed:
\begin{enumerate}
    \item Primordial matter is assumed to fall uniformly into each simulation sub-cube. The total amount of gas falling into the simulation volume is calculated via a \begin{equation}
\dot{ M}(t)= a  t^b  e^ {-t/\tau} \textit{,}
\end{equation}
prescription, which permits two main infall components: An initial constant rise of infall following by an exponential decay of the infall rate. While $\tau$ and the total Galaxy evolution time $t_{\mathrm{final}}$ are fixed initially, the parameters $a$ and $b$ can be determined alternatively from $M_{\mathrm{tot}}$ (the total infall mass integrated over time), defined by
\begin{equation} 
M_\mathrm{tot} = \int_0^{t_{\mathrm{final}}} a  t^b  e^ {-t/ \tau} dt\textit{,}
\end{equation}
 and the time of maximal infall $t_\mathrm{max}$, given by
\begin{equation}
t_\mathrm{max}=b \tau \textit{.}
\end{equation}
See table~\ref{infall parameters} for the applied parameters. 
\begin{table}
\begin{tabular}{|llr|}
\hline
\hline
$M_{tot}$ & Total infall mass & $10^8 \text{M}_{\odot}$ \\
$\tau$ & time scale of infall decline & $5\times 10^9$yrs \\
$t_{max}$ & time of the highest infall rate & $2\times 10^9$yrs \\
$t_{final}$ & duration of the simulation & $13.6\times 10^9$yrs \\
\hline
\end{tabular}
\caption{Main infall parameters. See Wehmeyer et al. (2015) for details on the parameters.}
\label{infall parameters}
\end{table}
\item The total gas mass in the volume is calculated and star formation is triggered. We use a Schmidt law with a density power $\alpha=1.5$ (since we aim star formation to be triggered by both the density of the ISM, as well as cloud-cloud interactions, see \citealt{Schmidt59,Kennicutt98,Larson91}) to determine to total mass of stars that are born in the current time step. This number is then divided by the integrated initial mass function (``IMF'', Salpeter type with a slope of $-2.35$) to obtain the number of stars formed per time step.
\item Once the number of newly born stars is calculated, star forming cells are selected randomly. Since star formation can be triggered by events as cloud-cloud interactions (e.g., shells of supernova remnants), we prefer cells with higher densities as location for newly born stars. 
\item Once a star forming cell is selected, we choose the mass of the newly born star randomly, with mass probabilities obeying a Salpeter type IMF with a slope of $-2.35$, in the mass range of $0.1 \text{M}_ {\odot} \leq m \leq 50 \text{M}_ {\odot}$\footnote{In this manuscript - when referring to stellar masses (excluding NSs and BHs) - we refer to the zero age main sequence mass of the star.}. In order to permit stellar masses to be well distributed (i.e., no bottom heavy IMF) we permit star formation only in cells containing at least $50 \text{M}_{\odot}$ of gas.  We consider stars with birth masses below $ 8 \text{M}_{\odot}$ as low and intermediate mass stars (LIMS), and stars more massive than $8 \text{M}_{\odot}$ as high mass stars (HMS)
\item The newly born star inherits the composition of the ISM 
out of which it is formed. From its birth mass and metallicity, we obtain its life expectation using the Geneva Stellar Evolution and Nucleosynthesis Group (cf. \citealt{Schaller92,Schaerer93a,Schaerer93b,Charbonnel93}) predictions, given by:
\begin{equation}
\begin{split}
\text{log}(t)&=(3.79 + 0.24  Z) - (3.10 + 0.35 Z)  \text{log} (M)\\
& + (0.74 + 0.11 Z)  \text{log}^2(M)\text{,}
\end{split}
\end{equation}
where $t$ is the expected life time of a star in My, $Z$ is the metallicity with respect to solar, and $M$ the star's mass in solar masses.
\item Once a star has reached the end of its calculated life time, a stellar death event is triggered (according to its birth mass), as discussed below.
\end{enumerate}

\subsection{Nucleosynthesis sites}
\subsubsection{Low and intermediate mass stars}\label{model:LIMS}

Low and intermediate mass stars (LIMSs) produce most of C and N in the Galactic disk \citep[e.g.,][]{Kobayashi11}. During the Asymptotic Giant Branch (AGB) phase, LIMSs produce the bulk of the slow neutron capture (``s-process'') abundances beyond Sr present in the solar system \citep[e.g.,][and references therein]{Kappeler11,Bisterzo14}.
LIMSs do not make significant contributions to the Galactic Fe or Eu inventory. Therefore, we only consider them as objects locking up gas for the duration of their life time for the purpose of our simulation. LIMSs return a significant amount of H and He in the ISM, marginally affecting the [Fe/H] ratios in the ISM. However, results and conclusions presented in this work are not affected. 
Once dying, LIMS give back portions of their locked up gas via stellar winds (resulting in a planetary nebula), leaving behind a white dwarf. Since planetary nebulae have observed sizes of a few tenths of to a few light years (e.g., Cat's eye nebula NGC 6543 with a 0.2 light year diameter, \citealt{Reed99}, Helix nebula with 2.87 light years, \citealt{Odell04}), for the purpose of our simulation, we simply return the locked up mass to the local cell once a LIMS has reached the end of its life time.

\subsubsection{Thermonuclear supernovae}\label{model:SNIa}
Since many stars in the Galaxy are born in double star systems (e.g., \citealt{Duchene13}), 
there is a chance that a newly born star has a companion that meets the prerequisites to let the double star system later undergo a supernova event of type Ia (SNIa). We follow the analytical suggestion of \cite{Greggio05} to simplify all associated stellar and binary evolution aspects to one probability ($P_{SNIa}=9 \times 10^{-4}$) for a newly born intermediate mass star (IMS, stars with masses in the range $1 \text{M}_\odot \leq m \leq 10 \text{M}_\odot$ ) to be born in a system that will later end up as a SNIa. 
This is equivalent to a rate of $7.49 \times 10^{-4}$ SNIa events per unit solar mass of stars formed. 
At the end of the life time of the second IMS, we inject $10^{51}$ erg of energy at the location of the event and emit the event specific yields (cf. \citealt{Iwamoto99}, model CDD2). As in \cite{Wehmeyer15}, we simply eject the same amount of Fe at all metallicities. This might be unrealistic (e.g., \citealt{Timmes03,Thielemann04,Travaglio05,Bravo10,Seitenzahl13,Leung18}), but this approximation does not strongly affect the outcomes of our simulation.
SNIa do not contribute to the r-process production, but they are the dominant source of Fe in the Galactic disk \citep[e.g.,][]{Matteucci86}. Therefore, we need to take into account the SNIa contribution to reproduce the chemical evolution of the [Eu/Fe] ratio in the Galaxy. 

\subsubsection{Core collapse supernovae and failed supernovae}\label{model:SNe}
Stars more massive than $10 \text{M}_{\odot}$ will experience all evolutionary stages until Si burning and the formation of an Fe core \citep[e.g.,][]{Jones13}.  
With the loss of its central energy source, the star cannot withstand the gravitational inward pull anymore and collapses. The core is compressed until it reaches nuclear densities, a so-called proto-NS. Neutrinos originating from the proto-NS lead to neutrino and anti-neutrino capture on neutrons and protons, which heat up matter in the so-called gain region (e.g., \citealt{Burrows13,Janka16,Janka17,Burrows18}) and lead to a successful explosion if the deposited energy is sufficient.  This is the case for a large fraction of initial stellar masses beyond $10 \text{M}_\odot$, but dependent on the stellar structure/compactness inherited from the pre-collapse stellar evolution this mechanism fails and results in the formation of a BH (e.g., \citealt{Heger03}).
In order to be able to determine when a star fails to explode instead of ending up in a supernova, the explosion energy predictions of the CCSN simulation suite PUSH (\citealt{Perego15,Curtis19,Ebinger19a}) are used to understand under which conditions massive stars collapse to a BH instead of exploding in a CCSN and leaving behind a NS. 
Their conclusions are that stars in the mass region $22.8 \text{M}_{\odot}\leq m \leq 25.6 \text{M}_{\odot}$ (at $\text{Z}= \text{Z}_\odot$) do not have sufficient explosion energies to withstand the gravitational collapse. These stars failing to explode in the CCSN simulations  are considered in the GCE suite in the following way: they collapse to a BH, without ejecting Fe.
Since most massive stars have at least one companion (e.g., \citealt{Duchene13}), we then use these results to constrain the BHNSM rate and the implications of this \textit{second} $r$-process site on the chemical evolution of the Galaxy (see section~\ref{subsec:NSM_BHNSM} for a detailed discussion of the implementation of BHNSMs/NSMs).

While we \textit{do} have prescriptions for the explodability and thus the production of metals by HMS at the end of their life time for \textit{solar} metallicity HMSs, it is expected that for low metallicities,  in contrast to to solar metallicities, a larger fraction of massive stars ends as BHs rather than CCSNe, due to smaller opacities and smaller amounts of mass loss during the hydrostatic phase.
Therefore, we employ the predictions made by these studies only close to solar metallicities and make different assumptions for lower metallicity HMSs: Since the explodability tends to scale with the progenitor compactness (\citealt{Ebinger17,Ebinger19a,Curtis19,Ebinger19b}), we employ the compactness of low metallicity progenitors at the time of the onset of the gravitational collapse as indicator whether the individual low metallicity progenitors will later undergo a successful CCSN. Lower opacity due to less metal content leads to less radiation scatter in the outer layers of lower metallicity stars.  This stellar wind loss has an effect on the compactness of stars: It leaves lower metallicity stellar cores at a higher compactness in comparison to their solar metallicity counterparts. Since the explosion calculations within the PUSH model have not yet been completed for lower metallicities, we utilize a simplified concept: In addition to the known explodability of solar metallicity HMSs, we test three extreme cases: all stars $\geq 20 \text{M}_{\odot}$ ($\geq 25 \text{M}_{\odot}$, $\geq 30 \text{M}_{\odot}$) at metallicities $\text{Z}\leq 10^{-2} \text{Z}_{\odot}$ (chosen to be metallicity-wise in between the current \citealt{Curtis19,Ebinger19a} predictions at Solar metallicity, and \citealt{Ebinger19b}, predictions for [Fe/H]$=-4$) are doomed to die in a failed SN at the end of their life time. This permits to account for the extent of the effects of the stellar wind mass losses, and therefore for the varied compactness of a lower metallicity HMS.

\subsubsection{NSM and BHNSM} \label{subsec:NSM_BHNSM}
If a double star system consists of two HMS, both end their life either in a failed or in a successful supernova \citep[e.g.,][]{Nomoto13}.
If the two remaining objects (two NSs in the case of two successful CCSNe, two BHs in the case of two failed SNe, and one NS and one BH in the case of one successful CCSN and one failed SN) survive the supernova kicks and remain gravitationally bound (e.g., \citealt{Tauris17}), this bound system emits gravitational waves and merges later. In this case, a compact binary merger (CBM, either a NSM or a BHNSM) event occurs.
BHNSMs can be an important source of r-process material. \citealt{Korobkin12} give results for the merger of a $1.4 \text{M}_{\odot}$ NS with either a $5\text{M}_{\odot}$ or a $10\text{M}_{\odot}$ BH, which produce comparable yield curves and ejecta masses to NSMs.
NSMs, on the other hand require two NSs and thus two successful CCSNe before the CBM event, so the surrounding ISM is already polluted with the ejecta of these two CCSNe and thus already enriched in metals. This means that the CBM products are ejected into a region where the metallicity is already high in comparison to the case of a BHNSM, where only \textit{one} NS is required, which means that only \textit{one} CCSN polluted the ISM with metals\footnote{Following this argumentation, BH - BH mergers might occur in a region where \textit{no} CCSN has occured and is thus metal-free. However, since BH - BH mergers do not eject any r-process enriched material, we do not consider this case here.}. 
Theoretical predictions for NSM rates vary strongly (e.g., \citealt{Kalogera04}), while the rates for BHNSMs are very controversial (e.g., \citealt{Mennekens14}).
Also, different nucleosynthesis (e.g., \citealt{Abbott1,Chornock17,Cowperthwaite17,Kasen17,Tanaka17,Wang17,Gompertz18,Hotokezaka18,Rosswog18}) and GCE studies (e.g., \citealt{Matteucci14,Cescutti15,Hirai15,Ishimaru15,Shen15,Wehmeyer15,Komiya16,Haynes19}) use different rates for this kind of event. \cite{Cote17} have compiled several modern GCE calculations involving NSM event probabilities and found that the rate assumptions differ by two orders of magnitude from study to study. This fact originates - among others - in the different treatment of infall prescriptions, differences in star forming prescription, employed IMF, CCSN/SNIa ejecta, and total ejected mass per NSM. When the assumptions in these studies are normalized to the same IMF, Fe yields, and Eu yields, then the number of NSMs per unit of stellar mass formed found in different studies converges within a factor of 4 (see \citealt{Cote17}).
While these theoretical prescriptions for NSM per unit volume or unit stellar mass formed vary greatly, a new approach helps us to determine the actual rate of CBMs in the local Universe: the detection of gravitational waves. While the first detections were attributed to BH - BH mergers (and are thus of less importance for this study) more recent ones have detected a NSM event (e.g., \citealt{Abbott4,Abbott5}, which predict a NSM rate of $1540^{+3200}_{-1220}$ Gpc$^{-3}$ yr$^{-1}$).
In order to reduce the number of free parameters in the formation channel, we use a simpler approach: We use an \textit{effective} probability factor $P_\text{r-proc}$, which represents the probability for a newly born HMS to be in a system that will end up as a NSM/BHNSM, producing the r-process.  We use $P_\text{r-proc}= 4\%$, which translates to (assuming a Salpeter initial mass function with a slope of $-1.35$, and a standard Cosmic star formation history with constant CBM delay times - see \citealt{Cote17} for the details of this conversion) $1.03 \times 10^{-4}$ CBM events per unit solar mass of stars formed. This rate is arguably high (see above and \citealt{Cote17} for a rate comparison of recent GCE models), but would correspond to an event rate of $\approx 1800$ Gpc$^{-3}$ yr$^{-1}$, which is well within the rate predicted by LIGO/Virgo.

However, this approach has one major caveat: If the BH in the binary system is too massive (or does not have sufficient angular momentum), this will lead to the NS either being swalled without disruption, or being disrupted and forming a disc, but inside the last stable orbit, i.e., not leading to mass ejection. The upper limit for BH masses to permit ejecta depends on the NS equation of state, the BH mass, and the BH spin (e.g., \citealt{Belczynski18}). With present knowledge (see \citealt{Rosswog15}), an upper limit seems to be in the range of $10$ to $14 \text{M}_\odot$ for the BH mass. Consequently, it is important how massive the resulting BHs would be after a star has undergone a failed SN. Two points need to be considered, (a) the mass loss during stellar evolution, and (b) which part of the pre-collapse star ends up in the BH and which part is still ejected in a failed SN. Possible options are that at least the H-envelope or all matter outside the CO core (or even more) is ejected. Looking at tables and figures in \cite{Thielemann18} and \cite{Ebinger19a}, referring to stellar models from \cite{Hirschi07,Limongi06a,Limongi06b} - and \cite{Woosley02,Woosley07}, respectively - with different rotation rates and metallicities, it turns out that for high (but credible, e.g., \citealt{Hirschi05}) rotation rates a $30 \text{M}_\odot$ star can loose half of its mass and an $80 \text{M}_\odot$ star can even end in a final pre-collapse mass of $20 \text{M}_\odot$. Including also the most recent results of \cite{Limongi18}, we find He-core masses below the above mentioned upper mass limit (for the disruption of the NS by the BH under ejection of r-process matter) for stars with initial masses below  $25$ to $30 \text{M}_\odot$, and CO-masses below these limits up to initial masses of $40 \text{M}_\odot$. Thus, while a point of caution should be kept in mind regarding the BHNSM scenario, it will clearly not be excluded. The occurrence rates utilized here should, however, be taken as an upper limit.

\section {Results}\label{Results}
\subsection{CBMs may explain r-process element GCE}
\begin{figure}
\includegraphics[width=\linewidth]{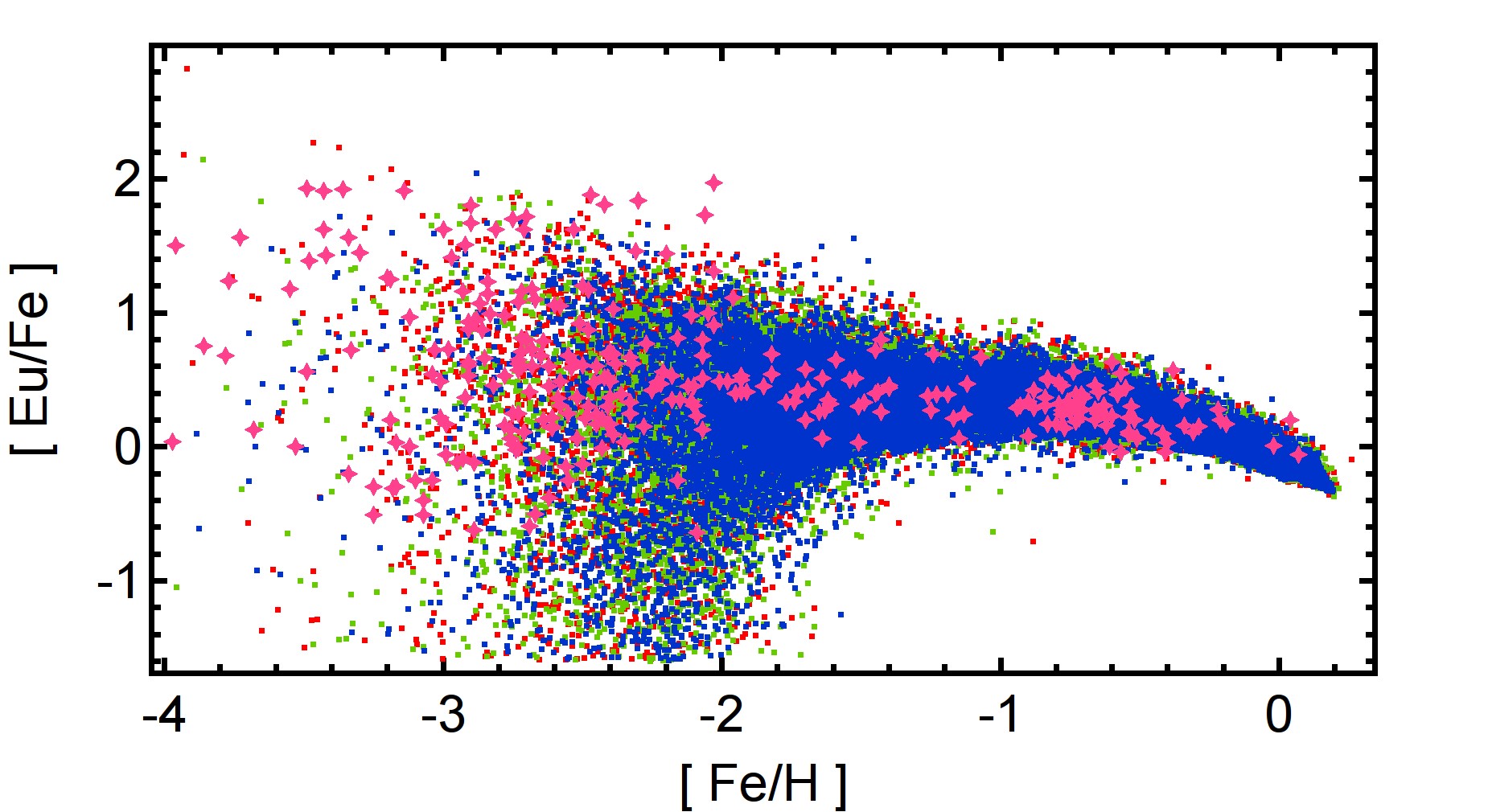}
\caption{Effect of the different choices of the prescriptions for failed SN at low metallicities on the GCE of [Eu/Fe]: Magenta crosses represent observations. Red (green, blue)  squares represent GCE models where all stars $\geq 20 \text{M}_{\odot}$ ($\geq 25 \text{M}_{\odot}$, $\geq 30 \text{M}_{\odot}$) at metallicites $\text{Z}\leq 10^{-2}\text{Z}_{\odot}$ are forming failed SNe at the end of their life.}
\label{metdep}
\end{figure}

\begin{figure}
\includegraphics[width=\linewidth]{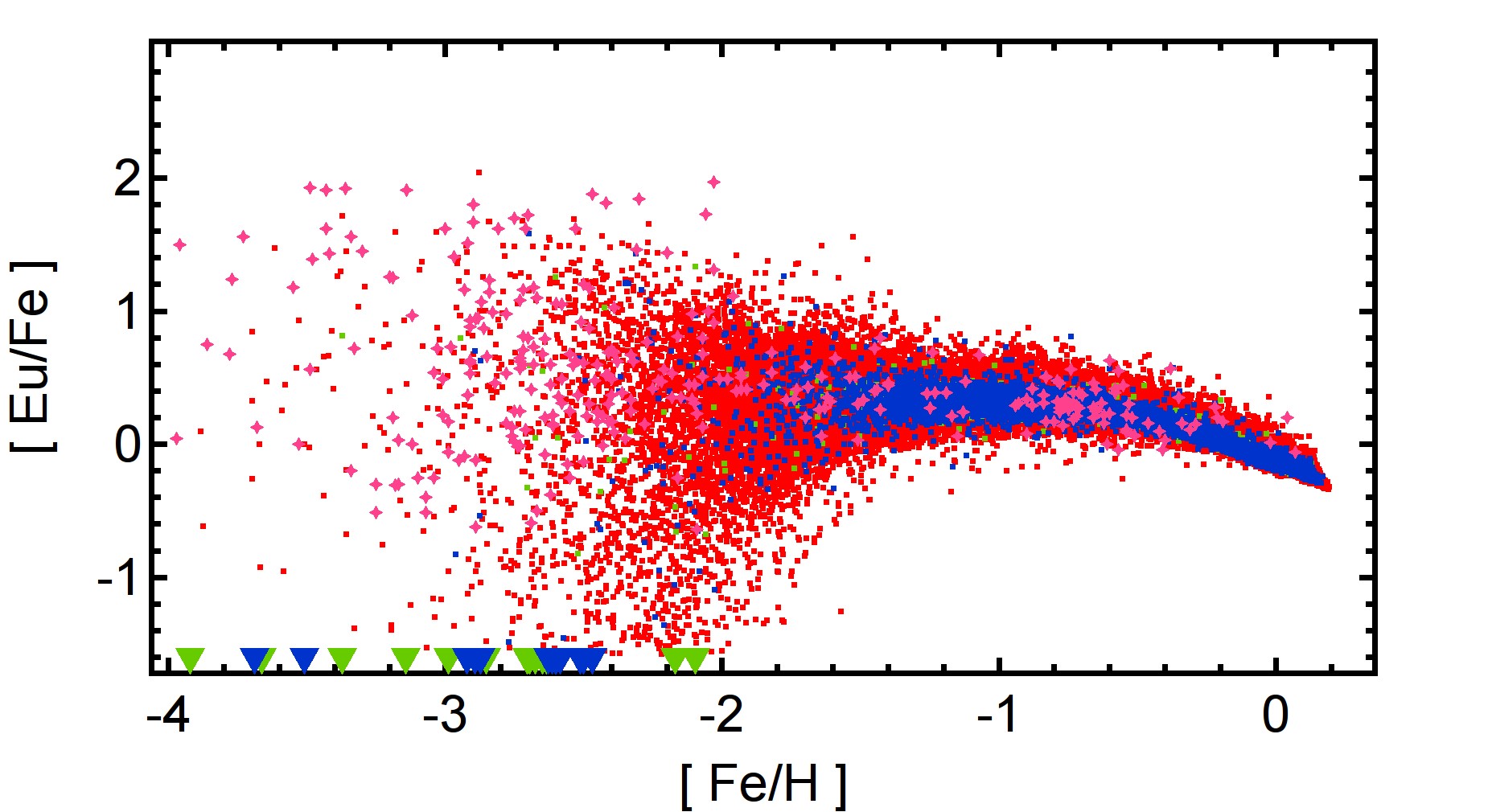}
\caption[]{Locations of NSM/BHNSM events in the [Eu/Fe] vs. [Fe/H] space of our fiducial model (failed SNe for $m \geq 30 \text{M}_{\odot}$ at metallicity lower than $\text{Z}\leq 10^{-2}\text{Z}_{\odot}$).  Magenta crosses represent observations. Red squares represent all model stars. Green and blue squares are the locations where BHNSMs or NSMs occur, respectively. This allows us to determine at what point the different r-process events contribute to the Galactic r-process element inventory.  Note that the first r-process events always have to occur in a r-process element free/poor environment, and thus are located at or near [Eu/Fe]$= -\infty$. We put green and blue triangles at the [Fe/H] locations above where the first BHNSM or NSM occur.
}
\label{metdeppos}
\end{figure}

Using our model, we study the effect of using BHNSMs as additional $r$-process production site. Our results suggest that the discussed deficiencies of using NSMs as exclusive $r$-process element production site can be cured by adding this second site. As can be seen in fig.~\ref{metdep}, both challenges mentioned in the introduction can be solved by using our model and including BHNSMs. Model stars (red, green, and blue squares) are \begin{enumerate}
    \item abundant in a very low metallicity region,
    \item show a large abundance scatter at lower metallicites, while this scatter is reduced towards higher metallicities, and
    \item are in qualitative agreement with the observations (magenta crosses)
\end{enumerate}  
This can be explained in the following way:
Regarding point \begin{enumerate}
    \item BHNSMs require only \textit{one} previous CCSN event (since they only require \textit{one} NS before the $r$-process event as opposed to \textit{two} previous CCSNe for NSM. This implies that this $r$-process event potentially happens at lower metallicities compared to NSM. See also section~\ref{subsec:NSM_BHNSM} for discussion, and fig.~\ref{metdeppos} for illustration.
    Additionally, another effect is relevant here: A model where a certain amount of stars fail to explode in a CCSN (and thus do not contribute to the Fe inventory of the Galaxy) slow down the [Fe/H] enrichment over time, compared to a model where every star succeeds to explode and thus contributes to the Fe evolution. This reduces the number of CCSNe per time step. See section~\ref{Age metallicity relation} for discussion.
    \item Since BHNSMs can occur while ejecting less Fe per $r$-process event (as discussed above), their event specific [Eu/Fe] (including the previous CCSN) is a factor of two higher in comparison two NSMs (where two CCSNe are required in order to form the two NSs). This potentially allows them to boost the abundances in terms of [Eu/Fe] much stronger than NSMs can. As can be seen in section~\ref{Dom_site}, the number of BHNSMs is higher in the beginning and subsequently lowers substantially. This leads to a decrease in the abundance boost and hence to less scatter in [Eu/Fe] abundances at higher metallicities.
    \end{enumerate}
Furthermore, if the mass range of failed supernovae in the IMF increases for lower metallicities, the event rate of BHNSMs increases accordingly and thus their nucleosynthetic influence towards low metallicities increases. This will be discussed in section~\ref{Dom_site}.

\subsection{Age metallicity relation}\label{Age metallicity relation}
\begin{figure}
\includegraphics[width=\linewidth]{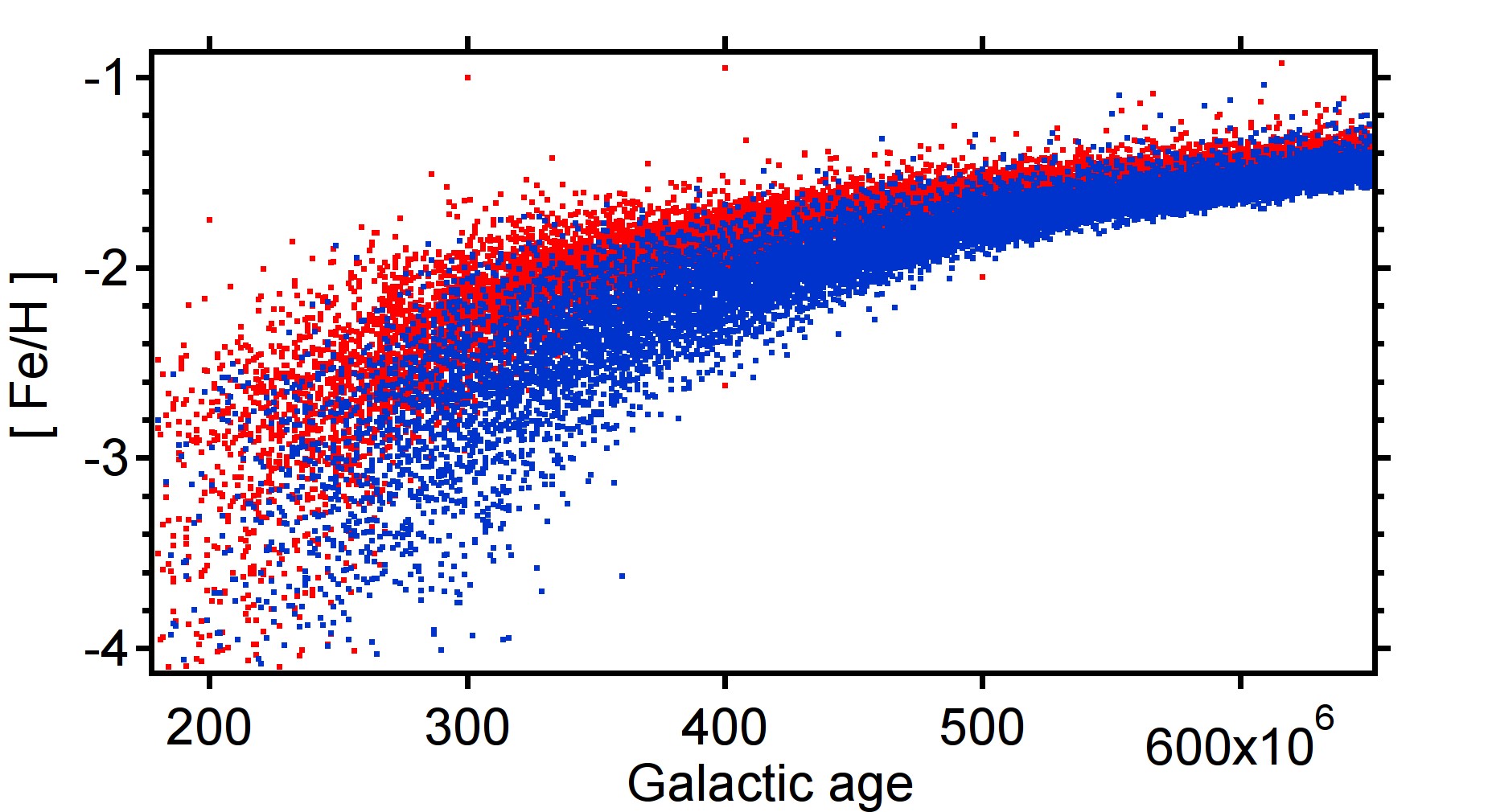}
\caption{Illustration of a shifted age-metallicity relation. Blue (red) squares represent model stars in a model that does (not) permit failed SNe. A model that permits failed SNe produces less Fe per time step, so the [Fe/H] enrichment is delayed in comparison to a model which does not allow failed SNe.}
\label{agezrel:plot}
\end{figure}

In a model where no failed SNe are allowed, all HMS die in a CCSN. So, all HMS eject Fe at the end of their life, and contribute it to the Galactic Fe inventory. Opposed to that, a model where failed SNe are allowed, some stars collapse into a BH. This means that those stars do not contribute to the Galactic Fe. If the same star formation rate for both of these models is assumed, a model permitting failed SNe has thus less CCSN events per time step compared to a model where all stars die in a CCSN. This leads to a slower increase in [Fe/H] vs. time. This also has implications on the GCE of r-process elements: All CBMs have a coalescence time between the death of the two stars and the merger event. When (in a model with enabled failed SNe) the [Fe/H] enrichment is slowed down with respect to time, the coalescence time scale of CBMs is of less importance. In other words, less nearby CCSN occur during the coalescence time. This allows CBM products to be ejected into a region that is less Fe rich than in a comparable model with no failed SNe permitted. See fig.~\ref{agezrel:plot} for illustration.
\subsection{The dominant r-process site}\label{Dom_site}
\begin{figure}
\includegraphics[width=\linewidth]{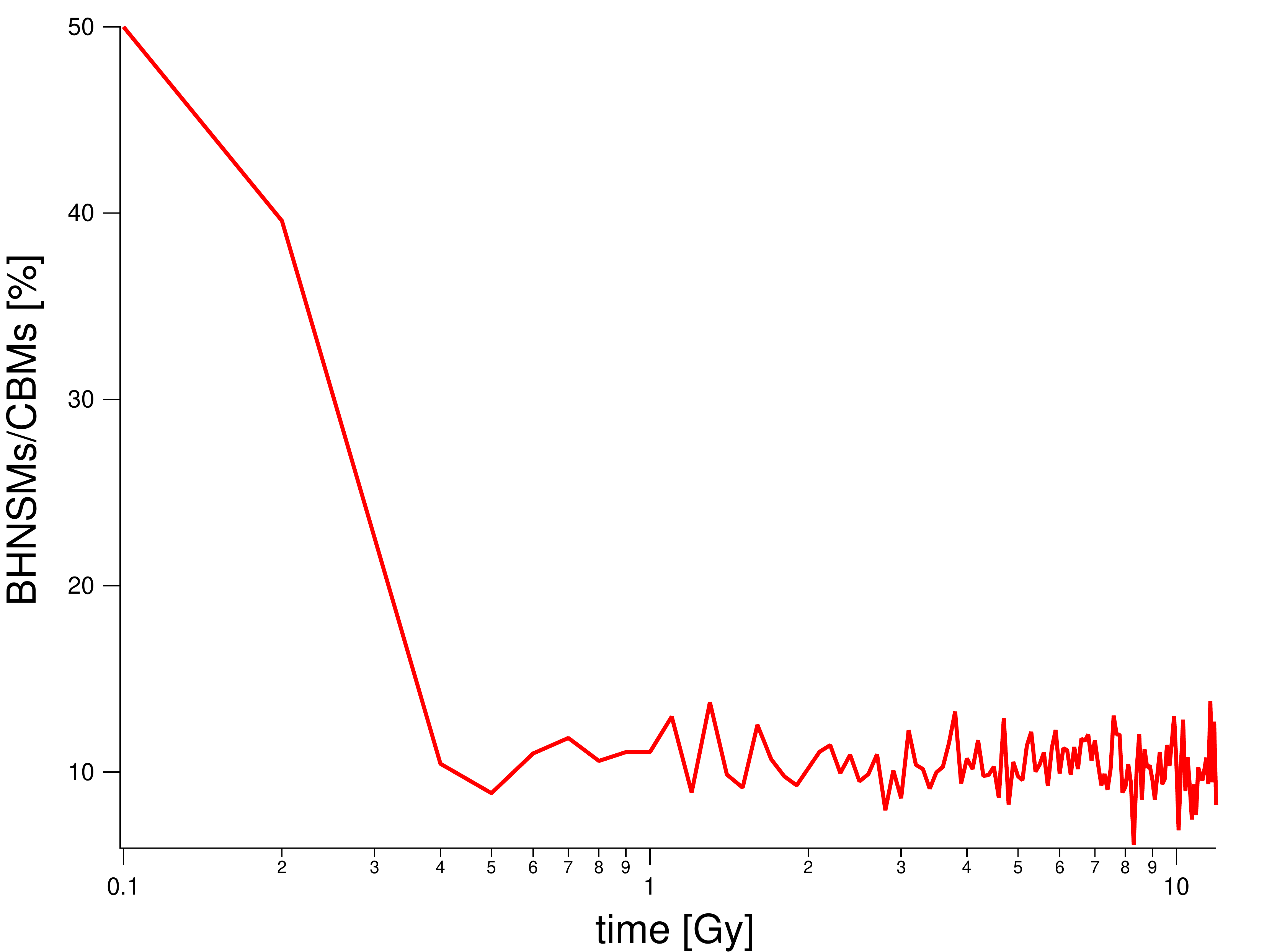}
\caption{Relative occurrence of BHNSMs with respect to all CBMs (BHNSMs+NSMs) using a model where stars $\geq 30 \text{M}_{\odot}$ at lower metallicity ($\text{Z}\leq 10^{-2}\text{Z}_{\odot}$), and stars $22.8 \text{M}_{\odot}\leq m \leq 25.6 \text{M}_{\odot}$ at higher metallicity die in a failed SN instead of a CCSN.}
\label{Dom_site:plot}
\end{figure}

Since in this simulation \textit{individual} stars and nucleosynthesis sites are followed, we can keep track of the number of individual events per time step. This allows us to determine which site (BHNSMs or NSM) is the \textit{dominant} site contributing to the r-process element production throughout the history of the Galaxy. Since NSMs seem to be the dominant site ($\geq 50\%$ of all CBM events at all times),  we consider the relative importance of BHNSMs with respect to overall CBMs ($=$BHNSMs+NSMs) in fig.~\ref{Dom_site:plot}. While the first r-process production events at early Galactic stages seem to be approximately equally performed by both types of CBMs, this changes rapidly towards NSMs as dominant r-process site. Already in early Galactic evolution stages ($t \geq 400$ My), the relative importance of BHNSMs in respect to all CBMs has reached its final value of $\approx 10\%$ of all CBMs. This originates in the fact that a large portion of stars (all stars $m \geq 30 \text{M}_\odot$) at lower metallicity ([Fe/H]$\leq -2$) will end up as a BH, while at higher metallicities only the stars in the range $22.8 \text{M}_{\odot}\leq m \leq 25.6 \text{M}_{\odot}$ will end up as BHs, according to the PUSH calculations utilized here.

\section {Conclusions and Discussion}\label{Conclusions and Discussion}
In this work, we have shown that the two major issues of the GCE of $r$-process elements, namely a) the large scatter in abundances in comparison to alpha-elements at lower metallicities, and b) that $r$-process elements are abundant at low metallicities, can be explained explained in our GCE model by including BHNSMs as a second $r$-process element production site in addition to NSMs.

 This scenario is complementary to magneto-rotational supernovae, or even collapsars, related to single stars and their early appearance in Galactic evolution, but the present study shows that BHNSMs could already produce the required effect. 

The main advantage of BHNSMs acting as a second $r$-process site is that, contrary to NSMs, only \textit{one} NS (plus one BH) is required to perform an $r$-process event. Hence only \textit{one}  previous successful CCSN is required, so the surrounding ISM is only polluted by Fe \textit{once} as opposed to \textit{twice} for NSMs. 
This advantage permits that BHNSMs occur in environments with less Fe content than the environment of NSMs. A second advantageous effect is that due to a higher failed SN rate at lower metallicities, i.e., less Fe-producing CCSNe, the overall enhancement of [Fe/H] is progressing slower in time, reducing the significance of the coalescence time scales of CBM.

Furthermore, we have shown that, despite that at early Galactic stages the r-process contribution of BHNSMs and NSMs to the Galactic $r$-process content is comparable, the contribution of NSMs is dominant over BHNSMs at later Galactic stages. This can be explained by more successful CCSN explosions with respect to failed SN explosion compared to lower metallicities, leading to a larger number of NSMs than BHNSMs.
\\
There remains a number of open questions in this work, related to the stochastic nature of this GCE approach (as already addressed in \citealt{Wehmeyer15}), as well as the specific implementation utilized this work.
\begin{enumerate}
\item We did not include CCSNe as $r$-process element sources, although there might be a chance for a small contribution to the abundance of r-process elements or a contribution to the ``weak'' r-process  by CCSNe. 
\item Also, we did not include the contribution of sub-halos (such as dwarf galaxies) to the chemical enrichment of the Galaxy.
\item Furthermore, we did not include magneto-rotational jet-supernovae or collapsars. They would have a similar, or even stronger (essentially emitting no Fe) effect, as described here for BHNSMs, but require strong assumptions on magnetic fields and stellar rotation, which would need to be confirmed observationally. 
\item The predicted rates for CBMs required to explain the chemical evolution are arguably high. They are well at the upper end of the spectrum in comparison to similar GCE calculations as inferred by \cite{Cote17}. Still, these are in overall agreement with the LIGO detection rates.
\item It has been shown by recent population synthesis studies (e.g., \citealt{Dominik12,Belczynski18,Chruslinska18}), that parameterized delay time \textit{distributions} (DTDs) should be used for CBMs instead of fixed coalescence time scales. Thus, our approach over-simplifies the GCE of $r$-process elements in the metallicity region of [Fe/H] $\geq -1$, omitting the modelling  difficulties associated with employing probably more realistic DTDs. See \cite{Cote17,Cote18}, and \cite{Hotokezaka18} for a discussion of this issue. A further effect, not yet considered here, could be that the coalescence time for massive binary systems containing one BH is possibly shorter than for NSMs.

\item Since the direct swallowing of a NS by a BH probably leaves no r-process matter behind, we did not consider this case here. Hence, our predicted r-process element production rate in section~\ref{subsec:NSM_BHNSM} omits this channel and thus has to be seen as a lower limit of a gravitational wave emission rate.
However, this event would not alter the conclusions of section~\ref{Age metallicity relation}, since the effect mentioned in that section originates only in the absence of Fe ejection by failed SNe (as opposed to successful Fe ejection in a case where \textit{all} CCSNe eject Fe).
\end{enumerate}
Future work towards the better understanding of the origin or the $r$-process elements will probably require
\begin{enumerate}
\item Detailed predictions of the explodability of low-metallicity stars being employed in a GCE model instead of a parametrized approach.
\item The efforts taken in this work should be re-examined using CCSN explodability predictions of different groups, e.g., \cite{Ugliano12,Ertl16,Sukhbold16}, and it should be examined whether this would change the required CBM event frequency, as well as the evolution of the BHNSMs/NSM ratio.
\item Future work should address the implications of CCSN kicks on the survival probability and dislocation of stellar binary systems (e.g., \citealt{Belczynski99}): If a kick by a CCSN was strong enough to make the binary system leave the supernova remnant bubble, the succeeding CBM event might take place in an area of the Galaxy that has not been polluted by CCSN ejecta before. Such an event might even contribute $r$-process elements at even lower metallicities than the CBMs happening inside a CCSN bubble considered in this work.
\item Future work should include DTDs instead of fixed coalescence time scales in this model. 
\item A future effort should be to include Jet-SNe as well as NSMs and BHNSMs as $r$-process element source. Of course, this would increase the level of complexity, since this would add another degree of freedom to the evolution of the Galaxy.
\item The next LIGO/Virgo run will probably provide us with a more accurate rate of CBMs. As soon as those are available, refined GCE calculations should be performed using these improved rates.
\end{enumerate}
 
\section*{Acknowledgements}
We thank Maria Lugaro, Chiaki Kobayashi, Albino Perego, Raphael Hirschi, Stephan Rosswog, Tim Beers, C. Gareth Few, and Brad K. Gibson for fruitful discussions.
We further extent our gratitude to Sanjana Curtis and Kevin Ebinger who have provided us with the PUSH predictions which play a crucial role for the considerations in this study.
\newline
BW is supported by a fellowship of the Swiss National Science Foundation (SNF).
BW and CF acknowledge support from the Research Corporation for Science Advancement through a Cottrell Scholar Award. CF was partially supported by the United States Department of Energy, Office of Science, Office of Nuclear Physics (award numbers SC0010263 and DE-FG02-02ER41216).
FKT was supported by the European Research Council (FP7) under ERC Advanced Grant Agreement No. 321263 - FISH, and the SNF. MP acknowledges the support from the SNF and the ``Lend\"ulet-2014'' Programme of the Hungarian Academy of Sciences (Hungary), of STFC, through the University of Hull Consolidated Grant ST/R000840/1, and access to {\sc viper}, the University of Hull High Performance Computing Facility. BW, BC, and MP acknowledge the support from the ERC Consolidator Grant (Hungary) funding scheme (project RADIOSTAR, G.A. n. 724560). BW, CF, BC, and MP acknowledge support of the National Science Foundation (USA) under grant No. PHY-1430152 (JINA Center for the Evolution of the Elements). This article is based upon work from the ``ChETEC'' COST Action (CA16117), supported by COST (European Cooperation in Science and Technology).

\end{document}